# Tunable nonlinear coherent perfect absorption with epsilon-near-zero plasmonic waveguides

YING LI AND CHRISTOS ARGYROPOULOS*

*Department of Electrical and Computer Engineering, University of Nebraska-Lincoln, Lincoln, NE, 68588, USA*
*Corresponding author:* christos.argyropoulos@unl.edu

**We propose a scheme to realize nonlinear coherent perfect absorption (CPA) at the nanoscale using epsilon-near-zero (ENZ) plasmonic waveguides. The general conditions to achieve CPA in a linear ENZ plasmonic waveguide are analyzed and presented. The proposed ENZ waveguides support an effective ENZ response at their cut-off frequency, where the CPA effect occurs under the illumination of two counter-propagating plane waves with equal amplitudes and appropriate phase distributions. In addition, the strong and uniform field enhancement inside the nanochannels of the waveguides at the ENZ resonance can efficiently boost Kerr nonlinearities, resulting in a new all-optical switching intensity-dependent CPA phenomenon which can be tunable with ultrafast speed. The proposed free-standing ENZ structures combine third-order nonlinear functionality with standing wave CPA interference effects in a nanoscale plasmonic configuration, thus, leading to a novel degree of tunable light-matter interactions achieved in subwavelength regions. Our findings provide a new platform to efficiently excite nonlinear phenomena at the nanoscale and design tunable coherent perfect absorbers.**

*(160.3918) Metamaterials; (190.0190) Nonlinear optics; (250.5403) Plasmonics.*

http://

Recently, increased interest has been dedicated towards the design of different plasmonic nanostructures to control light in an efficient and coherent way. One example is the effect of coherent perfect absorption (CPA) leading to the interferometric all-optical control of absorption under the illumination of two counter-propagating coherent beams [1]. As the time-reversed counterpart of lasing, coherent perfect absorption, also known as "anti-lasing", can be obtained by reversing the gain in a medium with absorption and completely annihilate the incoming beams due to the formation of a standing wave distribution inside a lossy medium [2]. This dynamic effect provided a new opportunity to dynamically modulate the absorption of a lossy medium by changing the relative phase between two counter-propagating incident waves. It has been investigated and demonstrated in various nanostructures [3,4] and new materials, such as graphene [5,6]. Moreover, CPA is also related to the new physics of parity-time (PT) symmetric photonic systems. Spatially symmetric and balanced inclusions of loss and gain in photonic structures have led to PT-symmetric devices that simultaneously behave as a coherent perfect absorber and a laser oscillator (i.e. a CPA-laser) [7,8]. However, CPA so far has been mostly realized with bulky materials exhibiting linear absorption and only a few theoretical studies have been focused on the extension of the CPA effect to nonlinear absorptive media [9–11].

In addition, during the recent years, considerable research efforts have been dedicated to realistic metamaterials with effective epsilon-near-zero (ENZ) permittivity response due to their peculiar transmission properties that provide, in principle, infinite phase velocity combined with enhanced and uniform field confinement distribution [12–14]. Recently, some interesting works have been proposed that connects ENZ metamaterials, even zero index metamaterials (ZIM), with CPA [15–17]. However, none of these works is focused on the nonlinear extension of the CPA effect.

In this Letter, we propose a nonlinear ENZ plasmonic waveguide system to realize tunable nonlinear CPA in nanoscale dimensions. We first analyze the CPA conditions in linear plasmonic waveguides. This plasmonic configuration exhibits an effective ENZ response at its cut-off frequency and Fabry-Pérot (FP) resonances at higher frequencies [18,19]. In the vicinity of the ENZ resonance, it is found that perfect CPA can be achieved under the illumination of two counter-propagating plane waves with appropriate amplitudes and phases. Next, we present an ultrafast and efficient way to switch ON or OFF the CPA process by incorporating Kerr nonlinear materials inside the ENZ plasmonic waveguide. Our findings can provide a new platform to excite nonlinear gap solitons [10], and design unidirectional CPA [20], optical switches [11], and ultrasensitive optical sensors.

The geometry of the proposed plasmonic waveguides' unit cell is shown in Fig. 1. It is composed of a narrow periodic rectangular slit carved in a silver screen whose permittivity dispersion follows derived experimental data [21]. The slits are loaded with a typical nonlinear dielectric material, which has a relative third-order nonlinear permittivity $\varepsilon_{ch} = \varepsilon_L + \chi^{(3)}|\mathbf{E}_{ch}|^2$, where $\varepsilon_L = 2.2$ represents the linear permittivity, $\chi^{(3)} = 4.4 \times 10^{-20}\,\mathrm{m}^2/\mathrm{V}^2$ is the third-order nonlinear susceptibility of silica [22], and $|\mathbf{E}_{ch}|$ is the magnitude of the local electric field inside the nanochannel. This free-standing waveguide geometry was originally introduced in [23] for other nonlinear applications and can sustain ENZ and FP resonances. The slit width $w$ is designed to tune the cut-off wavelength of the dominant quasi-TE$_{10}$ mode along the channel. The guided wave number $\beta$ of this mode has near-zero real part, resulting in effectively infinite guided wavelength and an anomalous impedance-matching phenomenon at the nanochannel. This effect is independent of the grating's periodicity and the waveguide channel's thickness [23]. Here, the slit dimensions of the rectangular channels are chosen to have width $w=200$ nm, height $t=40$ nm ($t \ll w$), and thickness $l=1\,\mu$m, respectively.

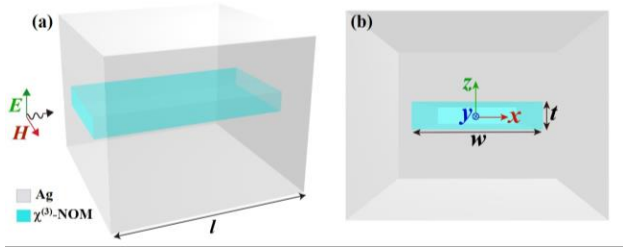

Fig. 1. Geometry of the silver plasmonic waveguide, where the rectangular slit is loaded with a nonlinear dielectric material ($\varepsilon_{ch} = \varepsilon_L + \chi^{(3)}|\mathbf{E}_{ch}|^2$). (a) The device is excited by a z-polarized plane wave impinging at normal incidence. (b) Cross-sectional view.

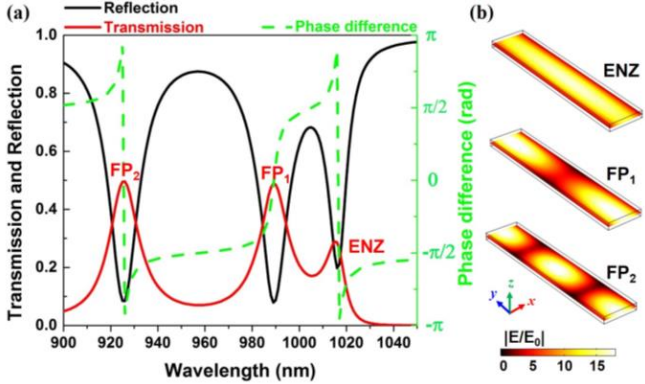

Fig. 2. (a) Transmission and reflection coefficients and phase difference between them for the plasmonic waveguide as a function of the incident wavelength. (b) The total electric field enhancement distribution in the channel's xy-plane operating at the ENZ, FP$_1$, and FP$_2$ resonant wavelengths.

First, we consider the linear operation of the plasmonic waveguide channel when the condition $\chi^{(3)}|\mathbf{E}_{ch}|^2 = 0$ is satisfied resulting to $\varepsilon_{ch} = \varepsilon_L = 2.2$. The linear plasmonic waveguide is now illuminated by a normal incident z-polarized plane wave and the simulated transmittance and reflectance are shown in Fig. 2 as a function of the incident wavelength. Resonant optical transmission peaks occur at the waveguide's cut-off wavelength and at lower wavelengths (red solid line in Fig. 2) [23]. At the cut-off wavelength of the dominant quasi-TE$_{10}$ mode ($\lambda=1016$ nm), the plasmonic waveguide behaves as an effective ENZ medium. An anomalous impedance-matching phenomenon occurs that leads to extraordinary transmission combined with large field enhancement and uniform electric field distribution inside each slit. The field distribution at the ENZ operation is shown in the upper caption of Fig. 2(b). Below the cut-off wavelength, the first and second order FP resonances are observed at $\lambda=989$ nm and $\lambda=926$ nm, respectively, which have typical standing wave field distributions [middle and lower captions in Fig. 2(b)]. The green dashed line in Fig. 2(a) refers to the computed phase difference $\Delta\phi$ between the transmission and reflection coefficients of the system. At the ENZ and lower FP resonances, the phase offset between the transmitted and reflected waves experience an abrupt transition from almost $\pi$ to $-\pi$, typical behavior of resonating effects. We will show later that either $\Delta\phi=0$ or $\Delta\phi=\pi$ is one of the required conditions to generate the CPA effect under the illumination of two coherent counter-propagating beams.

Next, we illuminate the same nanowaveguide with two plane wave sources of equal intensity $I_0$ ($I_0 = |\mathbf{E}_{in1}|^2/2\eta_0 = |\mathbf{E}_{in2}|^2/2\eta_0$) launched from opposite sides, as illustrated in Fig. 3(a). CPA occurs only when the transmission coefficient $t$ and reflection coefficient $r$ satisfy the condition $t^2 = r^2$ [1], which is valid for reciprocal systems. Hence, in order to achieve CPA through destructive interference, the transmission and reflection coefficients not only should have equal amplitudes ($|t|=|r|$), but they must also have a particular phase difference $\Delta\phi$, which is equal to either 0 or $\pm\pi$ [1]. All these conditions are very difficult to be simultaneously satisfied in subwavelength plasmonic structures and usually large photonic structures are used instead with several wavelengths thicknesses [24]. For instance, in our design, we consider a dielectric slab with the real part of permittivity $\varepsilon_R = 2.2$ and very small thickness $l=1\,\mu$m. According to Eq. (7) in [1], for a given incident wavelength (around 1$\mu$m), CPA would occur only when the imaginary part of permittivity $\varepsilon_I$ of this ultrathin slab will reach 0.7, an extremely high dielectric loss value that does not exist in nature. Moreover, just a metallic slab will also not work to obtain CPA because it will reflect all the incident radiation from both sides.

However, the aforementioned CPA conditions can be easily fulfilled by the proposed plasmonic waveguide system, which has a very small thickness ($l=1\,\mu$m) and is loaded with lossless dielectric material. As illustrated in Fig. 2(a), at the wavelength point ($\lambda=1017$ nm), very close to the ENZ wavelength, the amplitude of the transmission is totally equal to the reflection amplitude ($|t|=|r|$) and the phase difference $\Delta\phi$ between the transmission and reflection coefficients ($t$ and $r$) reaches nearly $-\pi$. Therefore, CPA is expected to be obtained at this point when we illuminate the nanowaveguide with two coherent counter-propagating plane waves ($\mathbf{E}_{in1} = \sqrt{2\eta_0 I_0}$ and $\mathbf{E}_{in2} = \sqrt{2\eta_0 I_0}\,e^{j\Delta\psi}$, where $\eta_0 = 377\,\Omega$ is the surrounding free space impedance) that have equal intensity amplitudes $I_0$ and zero phase offset $\Delta\psi=0$. To quantitatively measure CPA, we define an output coefficient $\Theta$ as the ratio of the total outgoing light intensity to the total incoming light intensity impinging at the nanowaveguide [7,25]: $\Theta = \left(|\mathbf{E}_{out1}|^2 + |\mathbf{E}_{out2}|^2\right)/\left(|\mathbf{E}_{in1}|^2 + |\mathbf{E}_{in2}|^2\right)$. Therefore, $\Theta = 0$ refers to CPA while $\Theta = 1$ refers to coherent perfect transmission (transparency). Figure 3(b) demonstrates the computed output coefficient $\Theta$ of the system as a function of the phase difference $\Delta\psi$ between the two counter-propagating incident waves ($\mathbf{E}_{in1}$ and $\mathbf{E}_{in2}$). The input wavelengths of the two incident beams are fixed at $\lambda=1017$ nm (the CPA point). Perfect CPA is obtained by using a phase offset $\Delta\psi$ value of either $0°$ or $360°$. The total output power can be modulated from zero (perfect CPA) to high transmission just by changing the phase offset $\Delta\psi$ between the two incident waves. Note that the high transmission obtained close to $\Delta\psi=180°$ is counterintuitive compared to the low transmission values obtained for just one incident wave shown in Fig. 2(a). Hence, the ENZ response becomes almost lossless in this case due to constructive interference and exhibits close to unitary transmission, an effect which is surprisingly achieved by a passive lossy plasmonic system. The normalized real part of the electric field distribution ($E_z$, the main component) at the CPA point ($\Delta\psi=0°$) is shown in Fig. 3(c), where the electric field outside of the plasmonic waveguide is nearly plus/minus unity (plane wave incidence), meaning that there are no reflections from either side and that all the incident radiation energy is completely absorbed by the proposed plasmonic waveguide (perfect CPA). This is another counterintuitive result, since the material at the core of the waveguide, where the electromagnetic energy is confined, is lossless. The white arrows in Fig. 3(c) represent the total power flow direction of each counter-propagating wave and provide another clear indication of the vanishing outgoing waves.

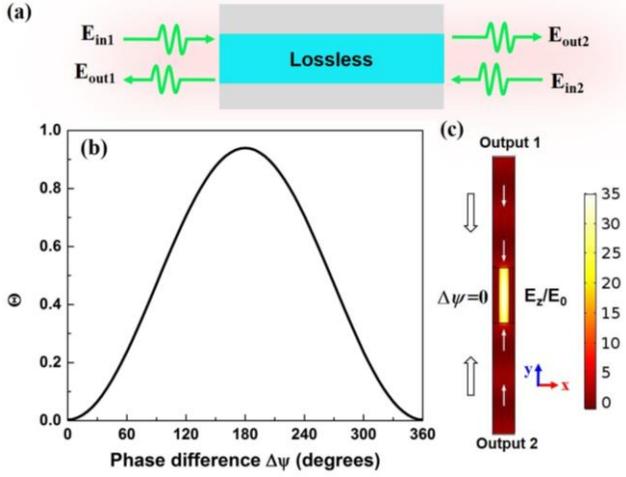

Fig. 3. (a) Plasmonic waveguide illuminated by two counter-propagating plane waves ($E_{in1}$, $E_{in2}$) from opposite sides. The two output plane waves ($E_{out1}$, $E_{out2}$) are also shown in the schematic. (b) Output coefficient $\Theta$ versus the phase difference $\Delta\psi$ of two incident waves operating very close to the ENZ wavelength. CPA occurs at $\Delta\psi=0°$ or $360°$. (c) Field enhancement distribution of the real part of $E_z$ in the channel's xy-plane at the ENZ wavelength in the case of $\Delta\psi=0°$ (CPA point). The white arrows depict the power flow direction.

In addition, it is interesting to note that the value of the field enhancement inside the nanochannels nearly doubles at the CPA point for the case of two incident waves, compared to the field enhancement at the ENZ resonance using only one incident wave [Fig. 2(b)]. However, the field distribution still remains homogeneous because the CPA effect happens close to the ENZ resonance of the waveguide. The obtained large and uniform field distribution at the CPA point is advantageous in order to increase light-matter interactions. For example, it will lead to large and uniform local and non-local density of optical states inside the waveguide's nanochannel, which are ideal conditions to increase the collective coherent spontaneous emission rate of several emitters placed inside the channel leading to superradiance [26]. In this case, CPA is achieved with an extremely thin plasmonic structure (thickness $l=1\mu m$) and, interestingly, the area of the field enhancement is lossless while the losses are coming only from its metallic walls.

The next step is to consider the Kerr nonlinear effect in the proposed CPA ENZ plasmonic waveguide. The strong and uniform field enhancement at the CPA point can cause the plasmonic system to access the optical nonlinear regime, especially as we increase the intensity of the incident waves. Third-order nonlinear effects will make the CPA phenomenon become intensity-dependent, in addition to phase-dependent, and, as a result, tunable with ultrafast speed. The nonlinear permittivity of the material loaded inside the waveguide channel was given before. The local electric field $\mathbf{E}_{ch}$ at the ENZ CPA point will be very strong and homogeneous [Fig. 3(c)] and the Kerr nonlinear effect is expected to be enhanced and triggered by relative low input intensities. This will directly lead to a shift in the frequency where CPA is obtained. Slow thermal nonlinear effects are not expected to affect the performance of the proposed nonlinear CPA device because it will be excited by relative low input intensities that will lead to low induced temperatures in the plasmonic structure.

Again, we illuminate the waveguide with two z-polarized plane waves [Fig. 3(a)] at the ENZ CPA point ($\lambda$=1017 nm) and compute the output coefficient $\Theta$ now versus the input intensity $I_0$ of each beam. For a given input intensity $I_0$, the phase difference $\Delta\psi$ of the two incident beams is set to vary from 0 to $2\pi$. We record the maximum (coherent perfect transmission) and minimum (CPA) values of $\Theta$ in Fig. 4(a). The CPA effect is always present [min($\Theta$) ~ 0] as we gradually increase the input intensity $I_0$ before reaching to a threshold value of 30MW/cm². Above this threshold low input intensity, min($\Theta$) gradually approaches max($\Theta$), which directly indicates that the nonlinear CPA effect disappears just by increasing the illuminating input intensities. Interestingly, an absorption dip in the value of $\Theta$ down to 0.0017 is observed at $I_0 = 3.16\,\text{MW}/\text{cm}^2$, attributed to the nearly perfect satisfaction of the CPA conditions at this point. It can be seen in Fig. 4(b) that for $I_0 = 0.01\,\text{MW}/\text{cm}^2$ (black line) CPA is obtained at $\Delta\psi=0°$ and $360°$ [similar to the linear operation in Fig. 3(b)], while at a larger input intensity $I_0 = 1000\,\text{MW}/\text{cm}^2$, the output coefficient becomes almost phase-insensitive [red line in Fig. 4(b)]. This can be explained by the fact that the third-order Kerr nonlinearity leads to a gradual change in the nonlinear permittivity of the dielectric medium and the system does not satisfy the aforementioned CPA conditions anymore. This effect is similar to nonlinear saturable absorption [22] but it works for waves illuminating the nonlinear structure from both sides and can lead to perfect transparency with relative low input intensities and in nanoscale dimensions. It can be used to achieve passive Q-switching for the envisioned compact nanolasers. It can also be used to design amplitude modulators for efficient all-optical routing of the electromagnetic radiation in nanoscale dimensions and to build adaptive optical filters. Such strong nonlinear tunable CPA response, achieved with relative low input intensities, is a unique feature of the proposed ENZ plasmonic structure. Note that regular elongated photonic CPA configurations will not be able to achieve this strong nonlinear response under such low input intensities due to the poor field confinement and enhancement inside their geometries.

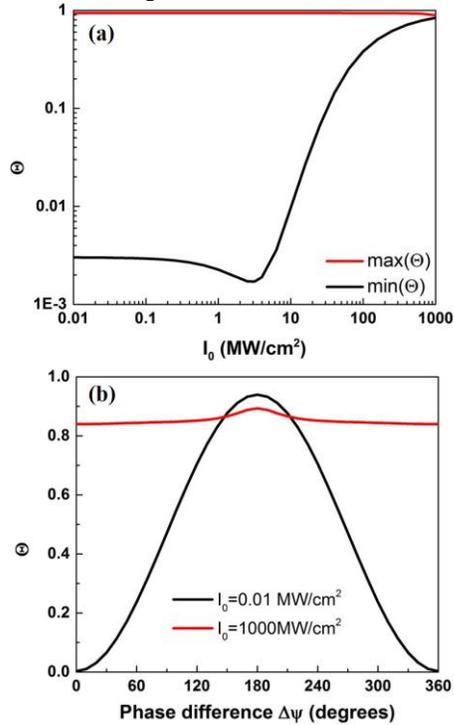

Fig. 4. (a) Nonlinear variation of the maximum and minimum values of the output coefficient $\Theta$ versus the input intensity $I_0$ for two incident counter-propagating waves operating very close to the ENZ wavelength. (b) The output coefficient $\Theta$ versus the phase difference $\Delta\psi$ for low 0.01 MW/cm$^2$ and high 1000 MW/cm$^2$ input intensity values.

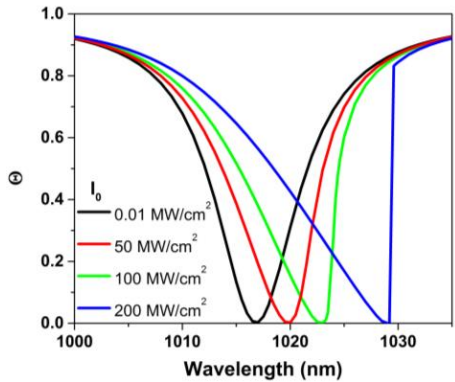

Fig. 5. Output coefficient $\Theta$ versus the incident wavelength for four different input intensities: $I_0$=0.01, 50, 100, and 200 MW/cm$^2$. The phase difference $\Delta\psi$ is set equal to zero leading to CPA at the ENZ point.

Next, we plot in Fig. 5 the nonlinear output coefficient $\Theta$ versus the incident wavelength around the ENZ resonance for four different input intensities. The phase difference $\Delta\psi$ of the two incident beams is now fixed and equal to zero, corresponding to perfect CPA. As we increase the input intensity $I_0$ from 0.01 MW/cm$^2$ to 200 MW/cm$^2$, the CPA effect always exists and the absorption dip (perfect CPA point) is gradually shifted to larger wavelengths and becomes tunable. Finally, it is interesting to note that the abrupt transition of $\Theta$ obtained for 200 MW/cm$^2$ is a clear indication of optical bistable response [23].

In conclusion, a nonlinear plasmonic waveguide with effective ENZ permittivity has been designed to achieve efficient tunable nonlinear CPA response at the nanoscale. At the ENZ resonance of the waveguide, the CPA conditions can be easily satisfied by the proposed plasmonic nanostructure. The strong and uniform field enhancement at the ENZ CPA point can efficiently boost third-order optical nonlinear effects and make the phase-dependent CPA phenomenon to also become intensity-dependent and, as a result, tunable with ultrafast (femtosecond scale) speed. As a final remark towards the practical implementation of the proposed structure, we stress that the ENZ plasmonic waveguide can be embedded in different dielectric materials that can be used as a substrate and superstrate without affecting the predicted CPA response. We believe that our results present a new platform to design low threshold all-optical switches, nanolasers, nonlinear gap solitons [10], and unidirectional coherent perfect absorbers [20].

**Funding.** National Science Foundation (NSF) (DMR-1709612) and Nebraska Materials Research Science and Engineering Center (MRSEC) (DMR-1420645).